\documentclass[reprint,twocolumn,superscriptaddress,amsmath,amssymb,aps,pre]{revtex4-1}
\usepackage{color}
\usepackage{graphicx}

\begin{document}
\title[]{Temperature anomalies of oscillating diffusion in ac-driven periodic systems}
%\title{Non-monotonic temperature dependence of oscillating diffusion\\ in ac-driven periodic systems}
%
\author{I. G. Marchenko}
\affiliation{NSC \lq\lq Kharkiv Institute of Physics and Technology\rq\rq, Kharkiv 61108, Ukraine}
\affiliation{Karazin Kharkiv National University, Kharkiv 61022, Ukraine}
\affiliation{Institute of Physics, University of Silesia, 41-500 Chorz{\'o}w, Poland}
%\author{A. Zhiglo}
%\affiliation{NSC \lq\lq Kharkov Institute of Physics and Technology\rq\rq, Kharkov 61108, Ukraine}
%\affiliation{Max Planck Institute for Astronomy, 69117 Heidelberg, Germany}

\author{V. Aksenova}
\affiliation{NSC \lq\lq Kharkiv Institute of Physics and Technology\rq\rq, Kharkiv 61108, Ukraine}
\affiliation{Karazin Kharkiv National University, Kharkiv 61022, Ukraine}

%\author{V. Tkachenko}
%\affiliation{NSC \lq\lq Kharkov Institute of Physics and Technology\rq\rq, Kharkov 61108, Ukraine}
%\affiliation{Kharkov National University, Kharkov 61077, Ukraine}

\author{I. I. Marchenko}
\affiliation{NTU \lq\lq Kharkiv Polytechnic Institute\rq\rq, Kharkiv 61002, Ukraine}

\author{J. {\L}uczka}
\author{J. Spiechowicz}
\affiliation{Institute of Physics, University of Silesia, 41-500 Chorz{\'o}w, Poland}

%\ead{j.spiechowicz@gmail.com}
%
\begin{abstract}
We analyse the impact of temperature on the diffusion coefficient  of an inertial Brownian particle moving in a symmetric periodic potential and driven by a symmetric time-periodic force. Recent studies  have revealed  the low friction regime in which the diffusion coefficient shows giant  damped quasi-periodic oscillations as a function of the amplitude of the time-periodic  force [I. G. Marchenko {\it at al.},  Chaos 32, 113106 (2022)]. We find out that when temperature grows  the diffusion coefficient increases at its minima, however, it decreases at the maxima within a finite temperature window. This curious behavior is explained in terms of the deterministic dynamics perturbed by thermal fluctuations and mean residence time of the particle in the locked and running trajectories. We demonstrate that temperature dependence of the diffusion coefficient can be accurately reconstructed from the stationary probability to occupy the running trajectories.   
\end{abstract}
\maketitle

\section{Introduction}
For the several last decades methods of diffusion processes have been applied in analysis of a very large number of phenomena not only in physics, chemistry, biology, material science or engineering but also in finance markets, communication, innovations \cite{rogers} and social systems \cite{socjology}. One of the most fundamental example of diffusion is Brownian motion, which is a universal phenomenon emerging both in classical and quantum world. For diffusion in classical systems and, in particular, for Brownian motion, it is commonly expected that an increase of temperature $T$ causes a growth of the diffusion coefficient $D$ \cite{entropy23}. This deeply held view is mainly based on the celebrated Einstein relation for which $D$ is a linearly increasing  function of $T$ \cite{risken}. It is also intuitive because for higher temperature thermal fluctuations are stronger and cause a larger spread of particle trajectories, which in consequence leads to an increase of $D$. 

However, there are systems which can exhibit the opposite effect, i.e., the diffusion coefficient decreases when temperature increases \cite{gang1996,roy2006,speer2012,marchenko2012,spiechowicz2015pre, spiechowicz2016njp,marchenko2017,marchenko2018,spiechowicz2015pre,spiechowicz2017chaos}. The necessary condition for such behavior is known: the system has to be in a nonequilibrium state. Sufficient conditions are not formulated in general. %The existing examples of non-monotonic dependence of $D$ on temperature $T$ do not allow to formulate general conclusions.  
One of the simpler systems which demonstrates  such an effect is a Brownian particle moving  in a spatially periodic potential under the influence of an external force. %The system studied in the paper operates in states far from equilibrium because 
Here we consider the case when the Brownian particle is subjected to a time-periodic force which drives it to a time dependent nonequilibrium state and in the long time limit it approaches a unique asymptotic state characterized by a temporally periodic probability density \cite{jung}. The diffusion coefficient, which describes the spread of trajectories and fluctuations around the average position of the particle,  depends on the parameters of the model, usually in a non-linear and sometimes in a non-monotonic way. An example is the above mentioned decrease of $D$ when temperature $T$ increases. 

Lately, another interesting non-monotonic behavior of $D$ has been found out, namely, the giant quasi-periodic changes of the diffusion coefficient when the amplitude of the time-periodic force increases \cite{march-chaos}. In such a case one can detect local maxima of $D$ separated by its local minima \cite{peter-epl,march-prl,renzoni1}. At the maxima, $D$ is much larger than the diffusion coefficient $D_0$ for the force-free thermal diffusion of the Brownian particle. At the minima, it is much smaller than $D_0$.  Here, we address the problem of the influence of temperature on this phenomenon, in particular when the diffusion coefficient attains its extremal values.

The paper is structured as follows. The presentation starts in Sec. II, where we define the system  in terms of the dimensionless Langevin equation of the Brownian particle moving in the spatially periodic potential driven by an unbiased time-periodic force. In Sec. III, we discuss the impact of temperature on the oscillating character of the diffusion coefficient $D$ and report regimes where the normal and abnormal temperature dependence of $D$ takes place. %It is the low damping case, which  accounts for inertial effects that play a crucial role in real experiments. 
Sec. IV contains the main points of analysis of the deterministic counterpart of the model which are crucial for noisy dynamics. The most important deterministic property is the existence of two classes of the particle trajectories, namely the running and localized ones. In Sec. V, we study the full noisy dynamics with emphasis on behavior of the diffusion coefficient. We consider regimes where its attains its extremal values. The impact of temperature on the diffusion coefficient is analyzed by applying several quantifiers like spread of trajectories, mean residence times in different states, etc. Sec. VI contains summary of the findings. In Appendix A, we detail the scaling scheme for Brownian dynamics and the corresponding dimensionless Langevin equation. Appendix B provides information about simulations of the Langevin equation. 

\section{Description of the model}
We consider the same model as in Ref. \cite{march-chaos}, which is formulated in
terms of the following dimensionless Langevin equation 
\begin{equation}
	\label{dimless-model}
	\ddot{x} + \gamma\dot{x} = -\sin{x} + a \sin (\omega t) +  \sqrt{2\gamma Q} \, \xi(t).
\end{equation}
In this scaling the dimensionless mass $m = 1$, the parameter $\gamma$ is the friction coefficient and $Q \propto k_B T$ is the dimensionless temperature of the system. The coupling of the particle with thermostat is modeled by the $\delta$-correlated Gaussian white noise $\xi(t)$ of vanishing mean, namely $\langle \xi(t) \rangle = 0$ and \mbox{$\langle \xi(t)\xi(s) \rangle = \delta(t-s)$}.
The starting dimensional equation is presented in Appendix A, where the corresponding scaling and dimensionless parameters are defined. The complexity of the underlying dynamics with the nonlinear force  $f(x) = -\sin{x}$, time-periodic force of the amplitude $a$ and frequency $\omega$, and thermal fluctuations of the intensity $2\gamma Q$ do not allow for a reasonable analytical approach which presently is clearly beyond known mathematical methods. Therefore numerical simulations were employed and their methodology is described in Appendix B. 

The Langevin equation (\ref{dimless-model}) and its quantum counterpart has been used for research in a wide range of mesoscale nonequilibrium phenomena. It describes driven transport in diverse physical systems including Josephson junctions \cite{kautz1996}, superionic conductors \cite{fulde} and cold atoms in optical lattices \cite{renzoni}, to mention only a few. Numerous diffusive  processes modeled by Eq. (\ref{dimless-model}) have been studied by many authors 
\cite{marchenko2012,spiechowicz2016njp,marchenko2017,spiechowicz2017chaos,bao,nash,kallunki,ikonen}  
and still new findings are emerging in this area. 

In Eq. (\ref{dimless-model}), the potential $V(x)=-\cos(x)$  corresponding to the conservative force $f(x)$ is symmetric and the time-periodic force $g(t)=a \sin{(\omega t)}$ is symmetric. Therefore in the long time limit the directed velocity $\langle v \rangle$ must vanish for both zero and non-zero temperature regimes \cite{denisov2014},  
\begin{equation}
	 \langle v \rangle = 0. 
\end{equation}
Here and below $\langle \cdot \rangle$ stands for the average over the ensemble of thermal noise realizations. Since the system is driven by the time-periodic force, the velocity  $\langle v \rangle$ can be calculated as
\begin{equation} 
	\langle v \rangle = \lim_{t \to \infty} \langle \mathsf{v}(t) \rangle,
\end{equation}
where
\begin{equation}
	\mathsf{v}(t) = \frac{1}{\mathsf{T}} \int_t^{t + \mathsf{T}} ds \, \dot{x}(s)
\end{equation}
is the particle velocity averaged over the period $\mathsf{T} = 2\pi/\omega$ of the external driving $a\sin{(\omega t)}$ \cite{jung}. Alternatively, the directed transport velocity $\langle v \rangle$ can be calculated by resourcing to a time averaged velocity
\begin{equation}
	\overline{v} = \lim_{t \to \infty} \frac{1}{t} \int_0^t ds \, \dot{x}(s), 
\end{equation}
which implies that
\begin{equation}
	\langle v \rangle = \langle \overline{v} \rangle. 
\end{equation}
In  the deterministic  case,  when $Q = 0$,  the dynamics  may be non-ergodic and therefore sensitive to the specific choice of the starting position $x(0)$ and  velocity $v(0)$ of the particle. In such a case we should modify the driving to the form $a\sin{(\omega t + \phi_0)}$, where $\phi_0$ is the initial phase.   Consequently, all quantities of interest should be averaged over $\{x(0), v(0), \phi_0\}$ with uniform distributions to get rid of this dependence. However, for any non-zero temperature $Q>0$ the system is ergodic and the initial conditions do not affect its properties in the long time stationary regime.
\begin{figure}[t]
\centering
\includegraphics[width=1.0\linewidth]{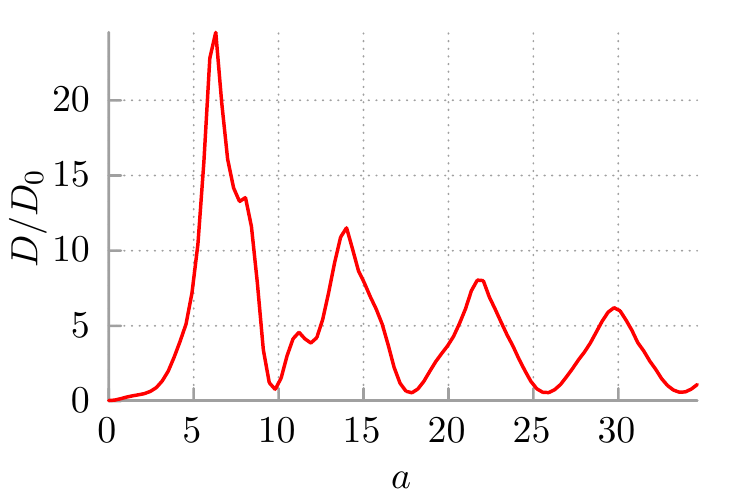}
\caption{The rescaled diffusion coefficient $D/D_0$, where \mbox{$D_0 = Q/\gamma$} is the force-free thermal diffusion coefficient, as a function of the amplitude $a$ of the ac-driving for the fixed value of its frequency $\omega=1.59$. Other parameters are: friction $\gamma = 0.03$ and temperature $Q=0.5$. } 
\label{fig1}
\end{figure}

\section{Diffusion anomalies}
The dynamical system described by the deterministic counterpart of Eq. (\ref{dimless-model}) exhibits an extremely rich behavior as a function of the four dimensionless parameters $\{\gamma, a, \omega, Q\}$. There are both regular and chaotic regimes, locked trajectories in which the motion is confined to a finite number of spatial periods and running states when it is is unbounded in space. 
A broad spectrum of locked and running states can be destabilized when random transitions between them are activated by thermal fluctuations at non-zero temperature. It potentially leads to a number of interesting phenomena \cite{gammaitoni1998, pikovsky1997, lindner2004, hanggi2009}. However, our goal is not to analyze all aspects of the system dynamics but we want to focus on diffusion properties in the selected parameter regime in which the diffusion process is normal in the long time limit and  is characterized by the diffusion coefficient \cite{spiechowicz2016scirep}
\begin{equation}
	\label{diffusioncoefficient}
	D =  \lim_{t \to \infty} \frac{1}{2t}\, \langle \left[x(t) - \langle x(t) \rangle \right]^2 \rangle.
\end{equation}
\begin{figure}[t]
\centering
\includegraphics[width=1.0\linewidth]{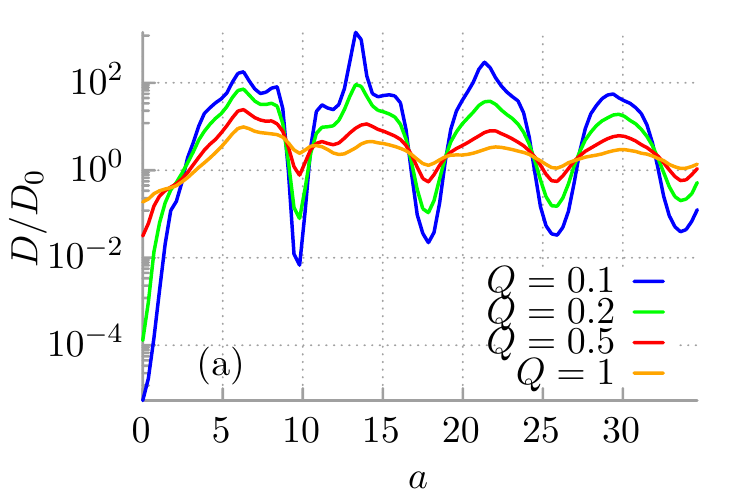}
\includegraphics[width=1.0\linewidth]{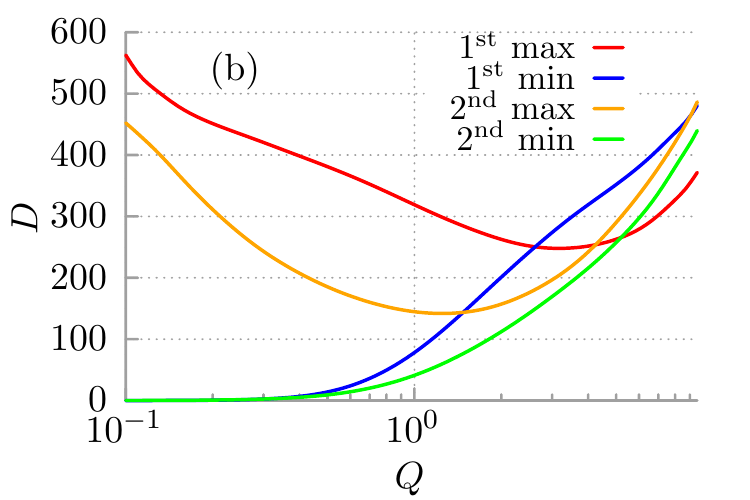}
\caption{The influence of temperature $Q$ on the diffusion coefficient. Parameters are the same as in Fig. \ref{fig1}. The first two maxima are for $a=6.3; 14$ and the first two minima are for 
$a=9.81; 17.85$.}
\label{fig2}
\end{figure} 
In Ref. \cite{march-chaos} it was shown that for a tailored parameter regime the diffusion coefficient $D$ displays the damped quasi-periodic dependence on the amplitude $a$ of the time-periodic force $g(t) = a\sin{(\omega t)}$. An example of such behavior is depicted in Fig. 1. The parameter regime in Fig. 1 is different than in Ref. \cite{march-chaos}, however, it displays the same qualitative features. We observe that indeed $D$ exhibits damped quasi-periodic oscillations as the amplitude $a$ increases. Moreover, there are local maximal values of $D$ which are much larger than the Einstein diffusion coefficient $D_0=Q/\gamma$. For example, if the driving amplitude is $a = 14$ and the dimensionless temperature is $Q=0.1$ the diffusion coefficient is $D \approx 1000 D_0$, see Fig. \ref{fig2}. On the other hand, at the local minima, $D$ is smaller than the Einstein diffusion coefficient. E.g., if the driving amplitude is $a \approx 9.81$ and temperature reads $Q=0.1$ the diffusion coefficient is $D \approx  0.01 D_0$. 
 
In this paper, we want to analyze the impact of temperature on the characteristics presented in Fig. 1. Typically, for higher temperature $Q$ the diffusion coefficient $D$ is expected to increase. However, we now know examples of systems for which the increase of temperature can reduce $D$ \cite{marchenko2012,spiechowicz2016njp,marchenko2017,spiechowicz2017chaos}. In Fig. 2(a) we show how  the quasi-periodicity of $D$ is deformed at four different values of temperature $Q$.  
One can notice three characteristic features: (i) there are intervals of the amplitude in the neighborhood of the local minima in which the diffusion coefficient $D$ increases when $Q$ grows (as expected); (ii) there are windows of $a$ near the maxima in which $D$ decreases when $Q$ increases (which is abnormal); (iii) there are values of $a$ at which $D$ is robust with respect to a change of $Q$, i.e. $D$ is very slowly varying function of $Q$. In other words, one can observe that when temperature is changed the diffusion coefficient is alternately reduced and boosted in adjacent intervals of the driving amplitude.  In Fig. 2(b) we display the diffusion coefficient for a larger window of temperature which includes high temperature limit when for any $a$ the diffusion coefficient increases. However, the most interesting is the case of abnormal dependence of $D$ on temperature. In the first maximum observed for $a\approx 6.3$, the abnormal diffusion emerges up to high temperatures $Q=4$, i.e. to $k_BT = 4 \Delta U$, see Eq. (A.7) in Appendix A. The deterministic counterpart of the studied dynamics is precursor of the above intriguing properties and therefore as the first step of our investigation we analyze the noiseless case.
\begin{figure*}[t]
\centering
\includegraphics[width=0.49\linewidth]{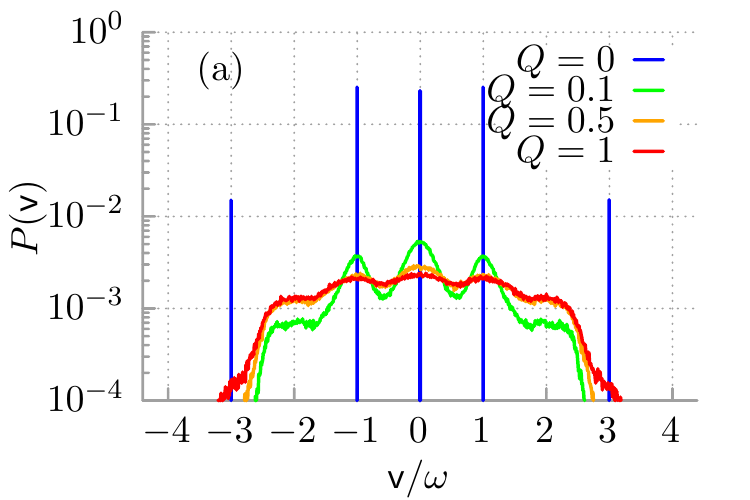}
\includegraphics[width=0.49\linewidth]{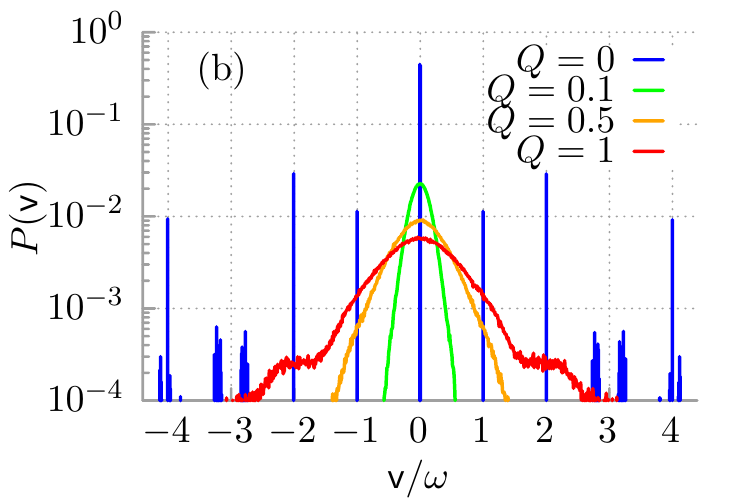}\\
\includegraphics[width=0.49\linewidth]{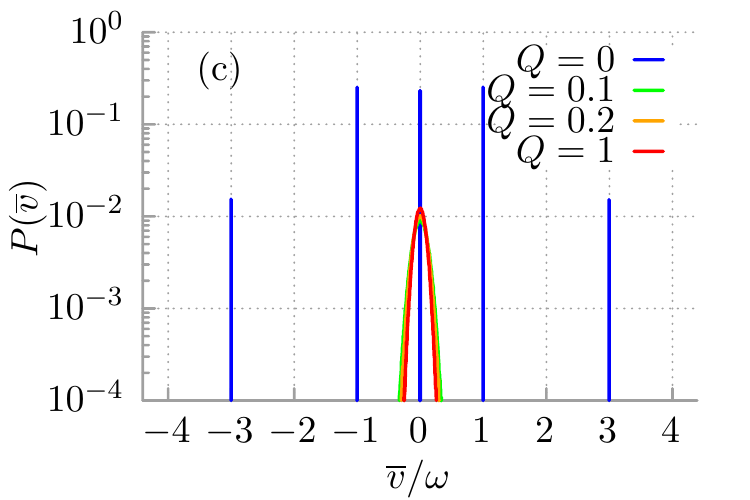}
\includegraphics[width=0.49\linewidth]{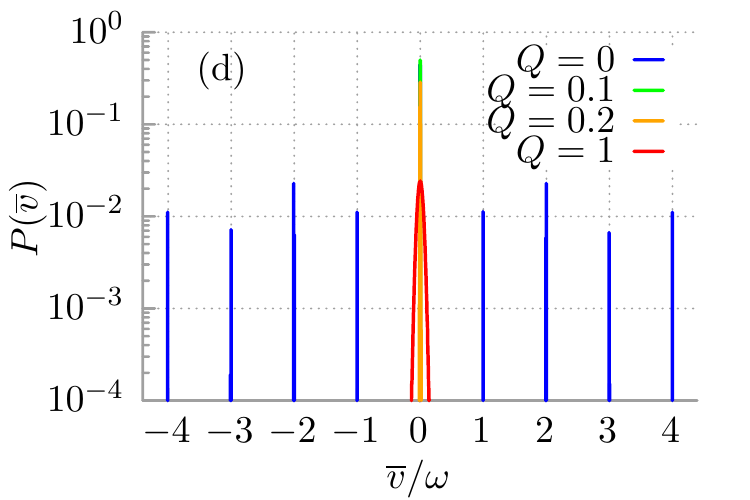}
\caption{Upper row: the probability distribution $P(\mathsf{v})$ for the period averaged velocity $\mathsf{v}$ of the Brownian particle in the long time regime is depicted for the first maximum ($a = 6.3$) and minimum ($a = 9.81$) of the diffusion coefficient $D$, c.f. Fig. \ref{fig2}, in panel (a) and (b), respectively. Bottom row: the probability distribution $P(\overline{v})$ for the time averaged velocity $\overline{v}$ is presented for $a = 6.3$ (panel (c)) and $a = 9.81$ (panel (d)). Other parameters are the same as in Fig. \ref{fig2}.}
\label{fig3}
\end{figure*} 

\section{Noiseless dynamics, $Q=0$}
The noiseless system corresponding to Eq. (\ref{dimless-model}) generally can exhibit both chaotic and non-chaotic behavior \cite{giter}.
%However, for the set of parameters as fixed in Fig. 1, the dynamics is non-chaotic and exhibits the various phase-space structures at maxima and minima.
For the parameter regimes corresponding to the maxima and minima of the diffusion coefficient $D$ both the running and localized states are detected. The particle can oscillate in the periodic potential wells and next proceed either forward or backward within one or many temporal periods, or it can permanently move in one direction, all depending on the initial conditions. Roughly speaking, at the maxima the population of both states is similar, however, at the minima the localized states are predominated over the running solutions. In Fig. \ref{fig3} the probability distribution of the system states is shown for two values of the driving amplitude $a=6.3$ (the first protruding maximum of $D$) and $a=9.81$ (the first protruding  minimum of $D$). Because both the potential and driving force is symmetric, the averaged velocity of the Brownian particle has to be zero, $\langle v \rangle = 0$, and consequently the distributions are symmetric. 

At the first  maximum $a=6.3$, there are five states  $\overline{v} \in \{\overline{v}_0=0, \  
\pm \overline{v}_{1} = \pm \omega, \ \pm \overline{v}_{ 3} = \pm 3\omega\}$, i.e., the localized state $\overline{v}_0$ (the particle oscillates in one or several potential wells and the motion is bounded in the coordinate space), two dominant running solution with the velocities $\pm \overline{v}_{1}$ and as well as two remaining running states $\pm \overline{v}_{3}$. One should pay attention to the fact that the vertical axis in Fig. \ref{fig3} is in the logarithmic scale and the population of the states with 
$\pm \overline{v}_{1}$ is much greater than the population of the states with $\pm \overline{v}_{3}$. At the first minimum $a=9.81$, there are nine relevant states $\overline{v} \in \{\overline{v}_0=0, \  \pm \overline{v}_{1} = \pm \omega,  \ \pm \overline{v}_{2} = \pm 2\omega, \ \pm \overline{v}_{3} = \pm 3\omega,  \ \pm \overline{v}_{4} = \pm 4\omega\}$, one dominant locked state $\overline{v}_0$, four running states with the positive velocities  as well as the same number of states transporting the particle in the negative direction.  
%Remark: because a limited set of initial conditions is used in simulations (see Appendix B), other running solutions  have not been detected. 
The structure of solutions is similar in the remaining maxima and minima presented in Fig. \ref{fig2}.

\begin{figure*}[t]
	\centering
	\includegraphics[width=0.49\linewidth]{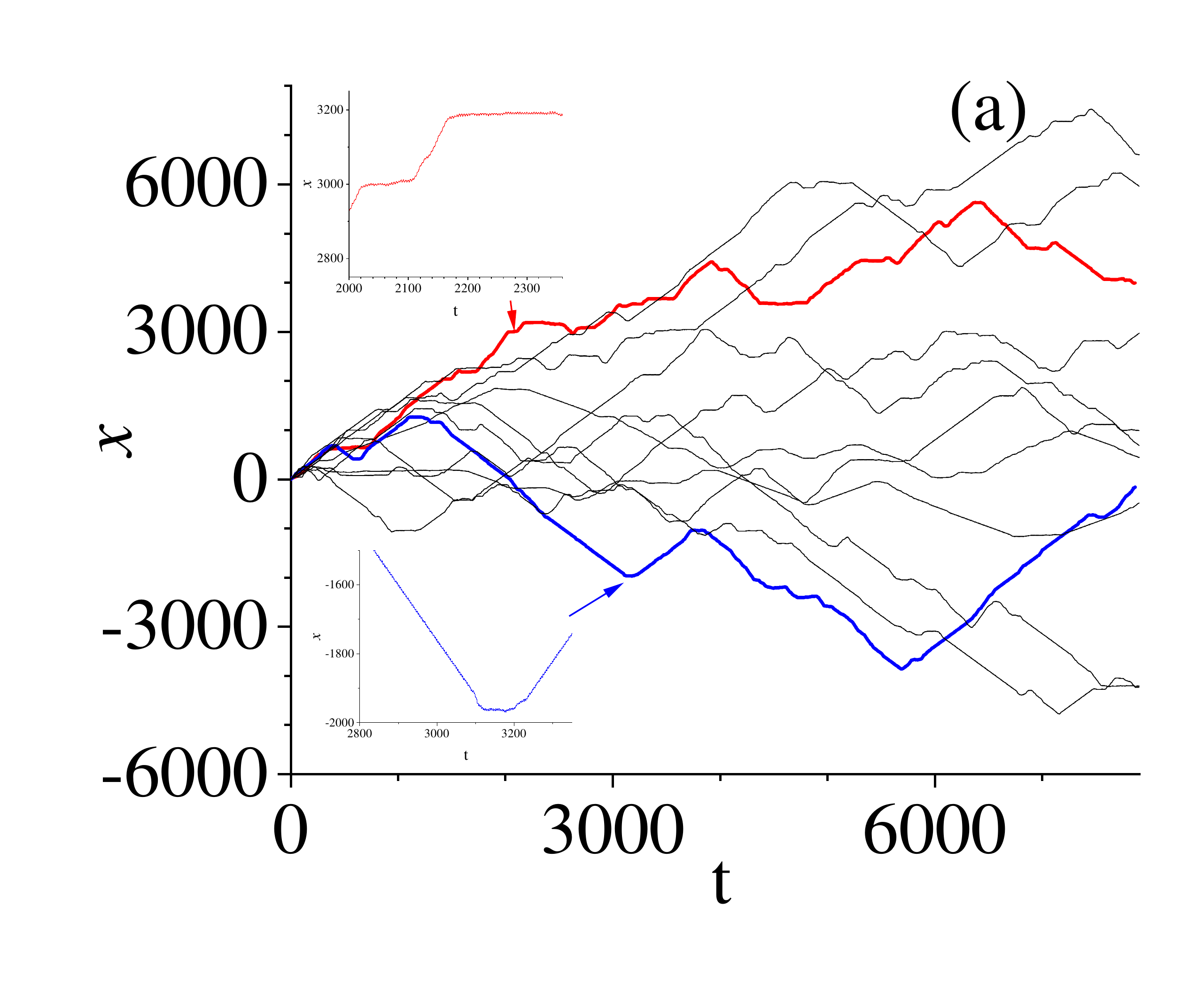}
	\includegraphics[width=0.49\linewidth]{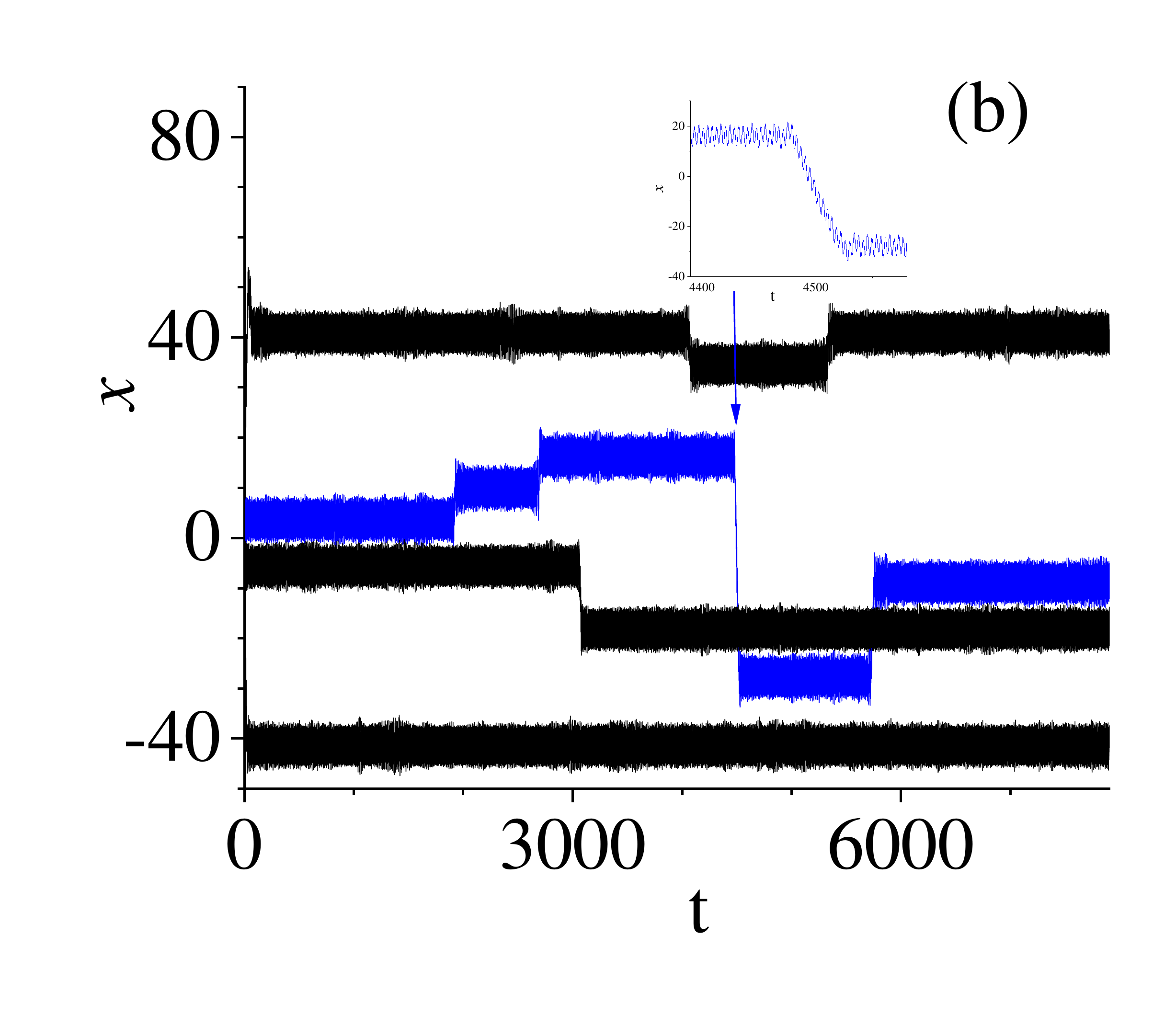}\\
	\includegraphics[width=0.49\linewidth]{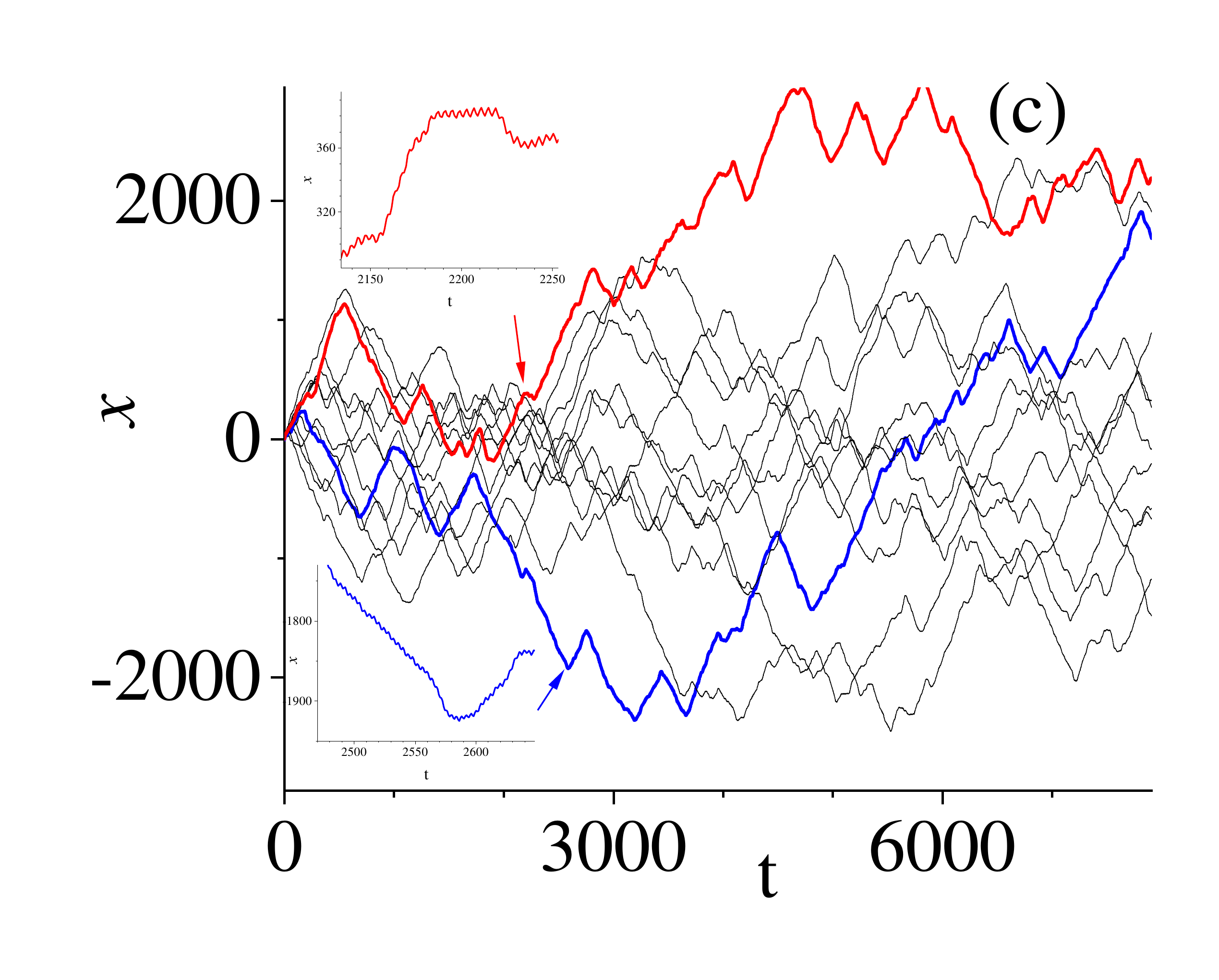}
	\includegraphics[width=0.49\linewidth]{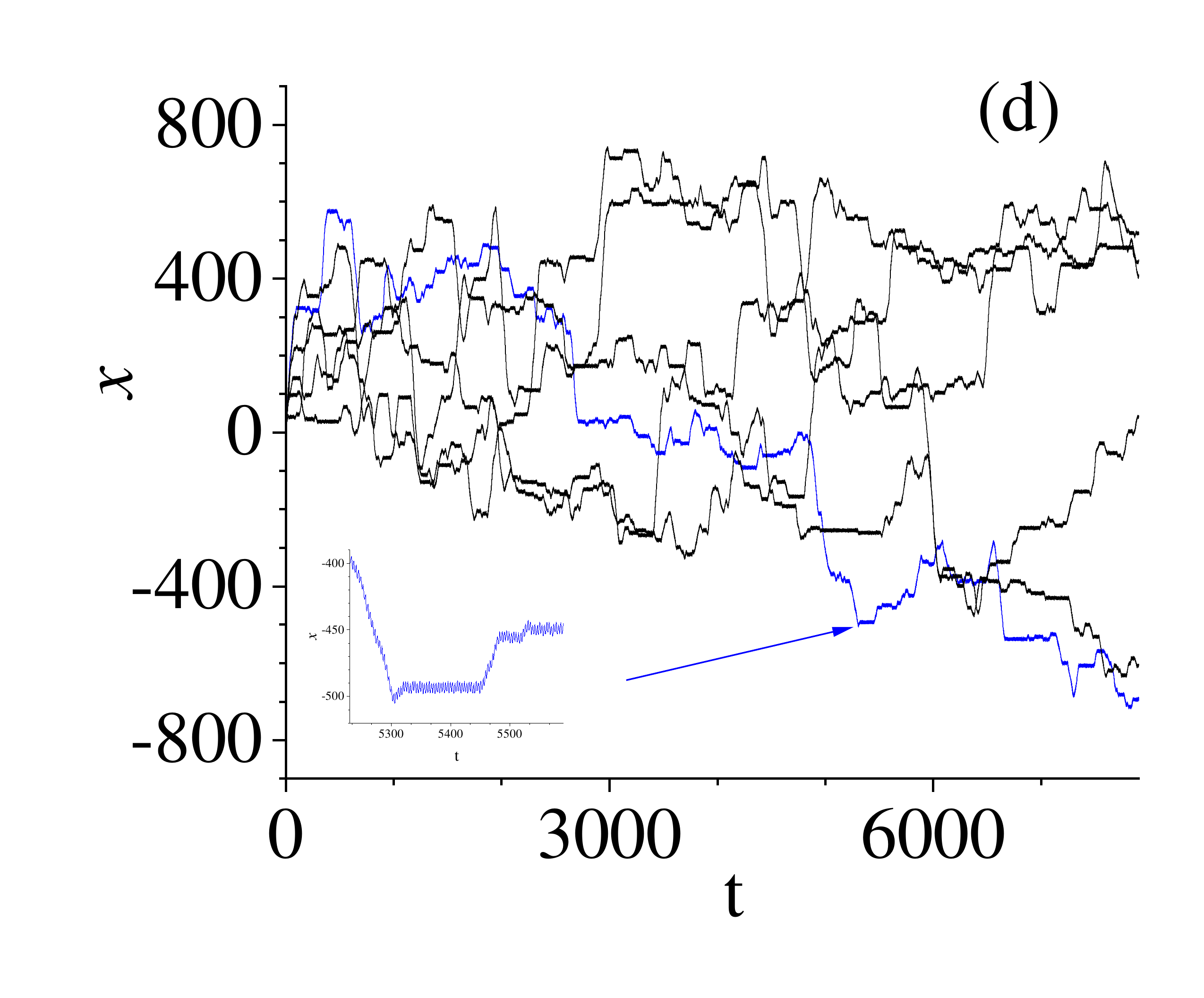}
	\caption{Exemplary set of the system trajectories at temperature $Q = 0.1$ (panels (a) and (b)) and $Q = 1$ (panels (c) and (d)). Left figures (a) and (c) correspond to the first maximum ($a = 6.3$) of the diffusion coefficient whereas the right (b) and (d)  present the first minimum ($a = 9.81$). Note the difference in the scale of ordinate axis. Other parameters are the same as in Fig. \ref{fig2}.}
	\label{fig4}
\end{figure*}     
\section{Noisy dynamics, $Q >0$}
Thermal fluctuations existing at non-zero temperature makes the noisy dynamics ergodic. Consequently, the initial conditions do not affect the results in the long time limit while their impact can be still relevant for transient regimes. The latter can last extremely long especially when the thermal noise strength $Q$ tends to zero. However, it is not the case for high temperature $Q \in [0.1, 1]$ presented in Fig. \ref{fig2},  where the asymptotic state is reached swiftly.

Although the asymptotic state is non-equilibrium and non-stationary, in some regimes it can effectively be described  in the framework  of equilibrium  statistical mechanics, at least for some observables. For example, in Ref. \cite{march-prl} the vibrational mechanics scheme is  applied and Eq. (1) is approximated by the Langevin equation of the type
\begin{equation}
	\label{dimless-model}
	\ddot{x} + \gamma\dot{x} = f(x) +  \sqrt{2\gamma Q} \, \xi(t),
\end{equation}
where $f(x)$ is some effective time-independent force. A similar approach is also presented in Ref. \cite{hindus}. Both approximations do not allow to explain the temperature dependence of the diffusion coefficient displayed in Fig. \ref{fig2}. Another approach is based on the concept of effective temperature  
$T^*$ which is a function of physical temperature $Q$ \cite{marchenko2012,santa}.  Then $D \sim T^*$ and   looks like the Einstein relation.  We have performed analysis and noticed that for very small temperature $Q < 0.05$ the agreement is quite satisfactory.  However, for the temperature regime $Q\in [0.1,10]$ considered in the paper we have observed a large deviation form this form. 
Therefore we have decided to apply numerical methods.

Let us consider diffusive spreading of the system trajectories. The particle starting from a given initial condition moves along the orbit according to dynamics determined by Eq. (1) and after some time it occasionally jumps onto a different solution. These random transitions are induced by thermal fluctuations and are realized as stochastic escape events connecting the coexisting attractors. In fact, the particle does not jump onto another trajectory but rather changes it in a continuous way. E.g. the particle can follow the running trajectory, next for some interval it resides in the localized state and later it again travels along the running trajectory either in the same or opposite direction. In this way, for a given trajectory various deterministic solutions are visited in a random sequence. We exemplify different scenarios of such behavior in Fig. \ref{fig4}. As a consequence the deterministic structure of locked and running states impacts the spread of trajectories. 

We note that there are several types of contribution to the spread of trajectories and as a result to the diffusion coefficient. The first, which is the largest one, comes from the distance between the running solutions moving into the opposite directions, e.g. $x(t) \sim \overline{v}_{1} t$ and 
$x(t) \sim -\overline{v}_{1} t$, that emerge in pairs due to the symmetry of the system. The second, also relatively large, is the spread between the running and localized trajectories, e.g. 
$x(t) \sim \pm \overline{v}_{1} t$ and $x(t) \sim x_k$ around the $k$-th well of the periodic potential. We note that as we presented above for a given parameter regime generally there are more than one running solution. Last but not least, there are contributions driven by thermal fluctuation in each of the running and locked states. Clearly, even at high temperature the latter are significantly smaller than the two previous ones. Therefore it is intuitively expected that the diffusion coefficient $D$ will change considerably especially when the population of running and locked states is significantly modified. In high temperature limit this subtle deterministic structure is blurred and looks like random motion of a free Brownian particle.

\begin{figure*}[t]
	\centering
	\includegraphics[width=0.49\linewidth]{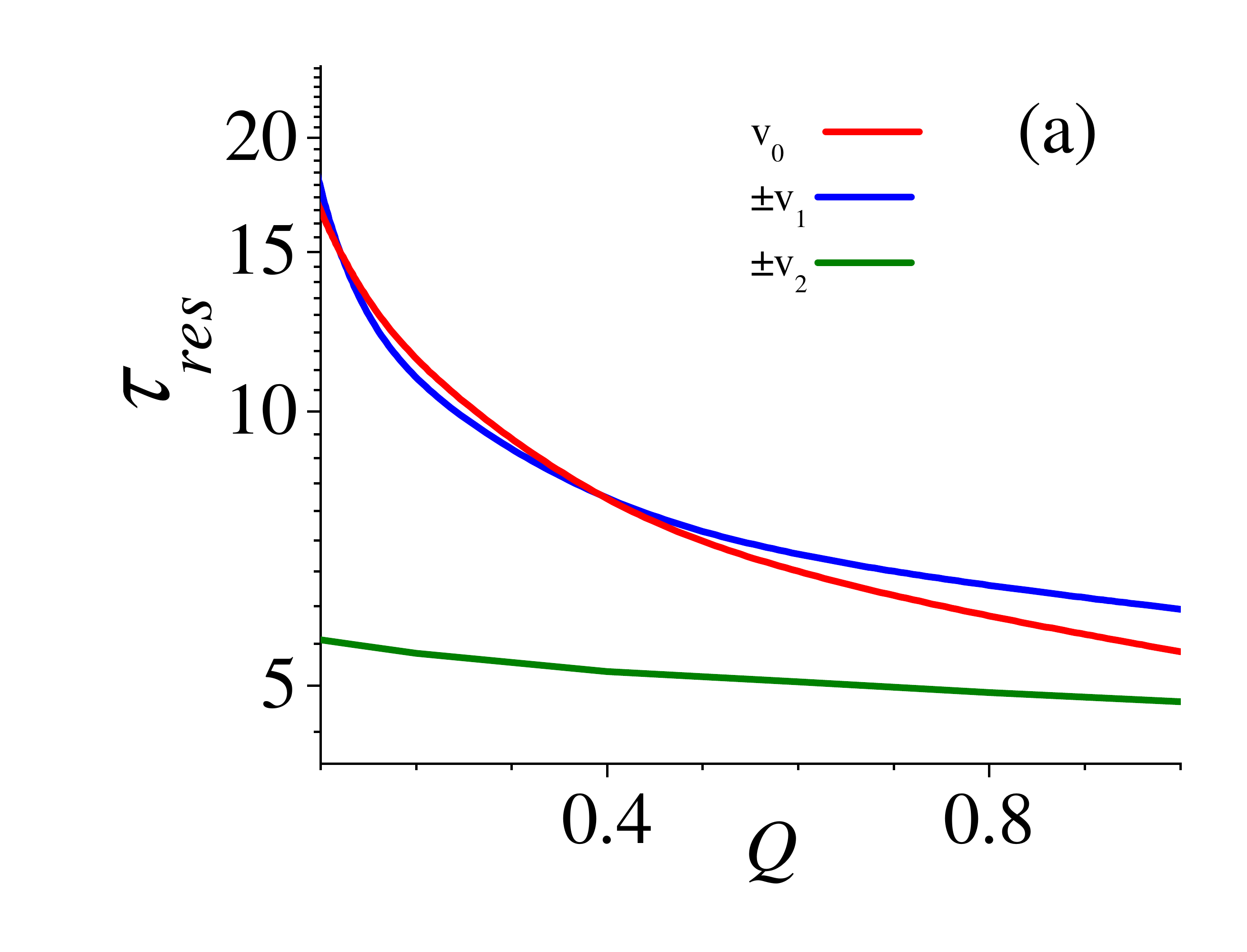}
	\includegraphics[width=0.49\linewidth]{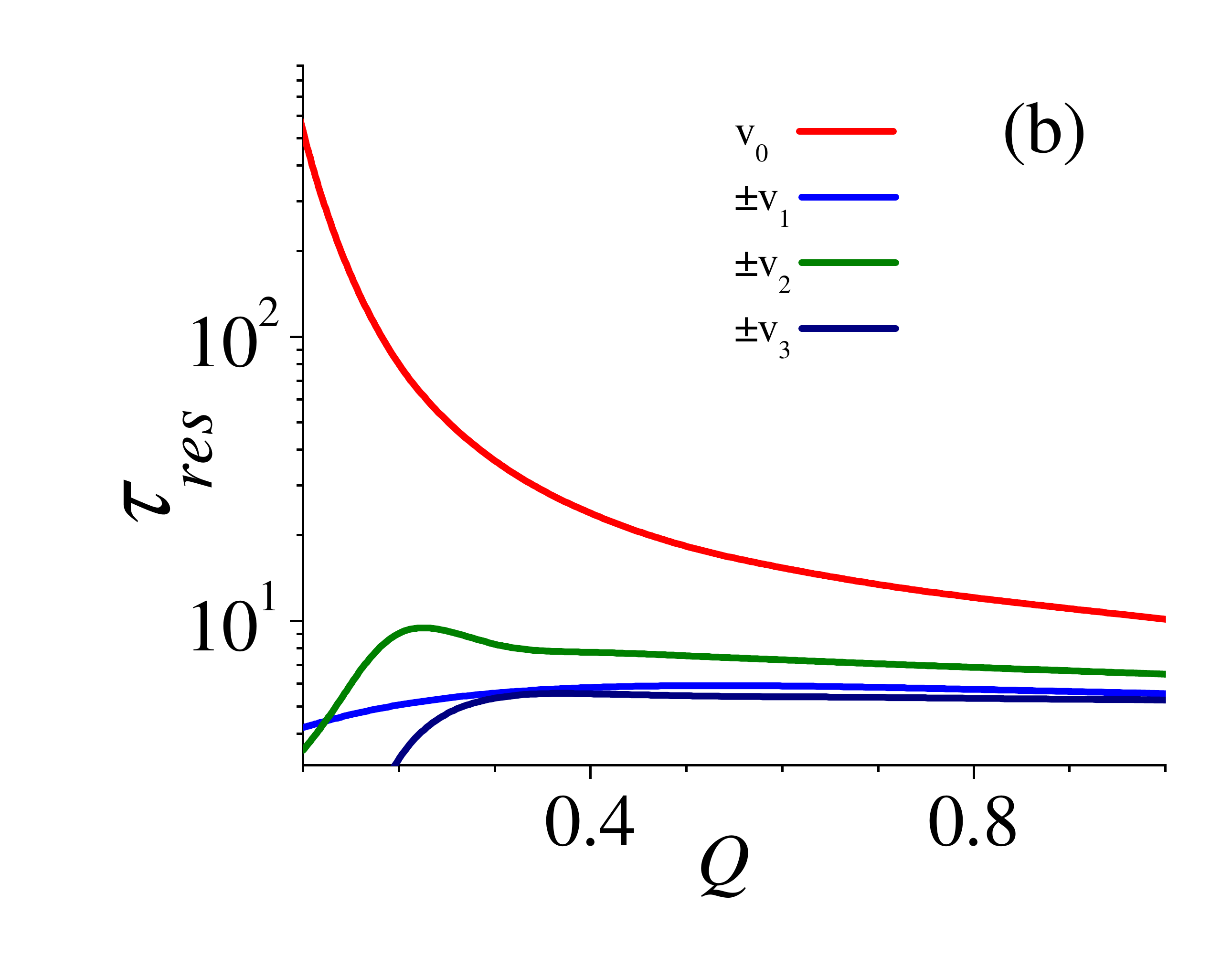}
	\caption{Temperature dependence of the mean residence time $\tau_{res}$ for different states is presented for the first maximum (panel (a), $a = 6.3$) and minimum (panel (b), $a = 9.81$) of diffusion coefficient. Other parameters are the same as in Fig. \ref{fig2}.}
	\label{fig5}
\end{figure*}
\subsection{Diffusion at maximum}
We continue our analysis by investigating temperature dependence of diffusion at the first maximum $a=6.3$. In the deterministic case, when $Q=0$, the system exhibits both the locked and running states $\overline{v} \in \{\overline{v}_0=0, \pm \overline{v}_1, \pm \overline{v}_3\}$, see panel (c) in Fig. \ref{fig3}. There are two dominant running solutions $\{\overline{v}_1, -\overline{v}_{1}\}$. The population of the locked states $\overline{v}_0$ is also large. There are also two running solutions $\{\overline{v}_3, -\overline{v}_3\}$ that are much less probable. When temperature increases in noisy system some states are eliminated in favor of the others (not depicted). E.g. for temperature $T = 0.01$ the system trajectories follow only $\overline{v} = \pm \overline{v}_1$ and the locked solution is completely wiped out. 

In panel (a) of Fig. \ref{fig4},  for temperature $Q=0.1$, one can observe the jagged piecewise locked  and running trajectories. The majority of the transporting parts have the slope $\pm \overline{v}_1$ and at first glance it is difficult to extract time intervals corresponding to the remaining solution $\pm \overline{v}_3$. It is related to the mean residence time $\tau_{res}$ in these states. We calculated this quantity from trajectories of the period averaged velocity $\mathsf{v}(t)$ simply by counting how many periods the particle stays in a given state until it jumps onto the other one. We observe the difference between the probability distributions $P(\mathsf{v})$ and $P(\overline{v})$ for the period averaged  and time averaged  velocities, respectively; see Fig. \ref{fig3} (a) vs (c). In the latter for non-zero temperature $Q \neq 0$ it is difficult to distinguish individual running states while for the period averaged velocity even for noisy system one can clearly observe trace of the deterministic structure of solutions $\mathsf{v} \in \{\mathsf{v}_0=0, \pm \mathsf{v}_1 = \pm \omega\}$. Moreover, it turns out that thermal fluctuations induce new running state $\pm \mathsf{v}_2 = \pm 2 \omega$ which is not present for the deterministic system.

In Fig. \ref{fig5} (a) we present temperature dependence of the mean residence time $\tau_{res}$ for  the noisy  states $\mathsf{v} \in \{\mathsf{v}_0=0, \pm \mathsf{v}_1 = \pm \omega, \pm \mathsf{v}_2 = \pm 2 \omega\}$  determined by the period averaged velocity $\mathsf{v}$.  For temperature $Q=0.1$ the corresponding mean residence time is $\tau_0 = 16.53,  \tau_{\pm 1} = 16.07, \tau_{\pm 2} = 5.43$. Therefore the marginal states $\mathsf{v}_{\pm 2}$ are rarely occupied. The main contribution for the swift spreading of trajectories and consequently to the large diffusion coefficient comes from long residence time $\tau_{\pm 1}$ corresponding to solutions that transport the particle in the opposite directions. If temperature decreases the mean residence time $\tau_{\pm 1}$ rapidly grows and dominates the other states. On the other hand, when temperature is high enough the difference between the average time which the particle spends in each of the states is marginalized. E.g. for temperature $Q = 1$ the residence times are $\tau_0 = 5.45,  \tau_{\pm 1} = 6.4, \tau_{\pm 2} = 4.9$. As it is shown in Fig. \ref{fig2} (b) for the first maximum $a = 6.3$ the diffusion coefficient $D$ for higher temperature $Q = 1$ is almost twice as small as for lower temperature $Q = 0.1$. The main reason for this behavior is the mentioned rapid growth of the mean residence time $\tau_{\pm 1}$ in the state $\mathsf{v}_1 = \pm \omega$ as temperature drops down. Conversely, if the thermal noise intensity increases the population of running states with long residence time is significantly reduced. It means that there are many transporting orbits which are  short-lived and therefore cannot generate such a large spread as for lower thermal fluctuation intensity. In the high temperature limit the periodic potential as well as the external driving can be neglected and this regime imitates a free Brownian particle.

\subsection{Diffusion at minimum} 
As the next point of our analysis we consider diffusion for the driving amplitude $a=9.81$ corresponding to the first local minimum, see Fig. \ref{fig2}. The set of deterministic solution encompasses nine states $\overline{v} \in \{\overline{v}_0=0, \pm \overline{v}_1, \pm \overline{v}_2, \pm \overline{v}_3, \pm \overline{v}_4\}$ in which the locked one $\overline{v}_0 = 0$ is dominant, c.f. panel (b) of Fig. \ref{fig3}. It is important to note that as temperature grows the running states are completely destroyed and only $\overline{v}_0 = 0$ survives. It must be contrasted with the case of diffusion at the maximum for $a = 6.3$ where temperature induces new states that are missing in the deterministic dynamics. In Fig. \ref{fig5} (b) we present the mean residence time $\tau_{res}$ for the selected noisy  states determined by the period averaged velocity $\mathsf{v}$. 
%We note that for the parameter regime corresponding to minimum of the diffusion coefficient the distinction between the locked solution $\mathsf{v}_0 = 0$ and the running states $\mathsf{v} \neq 0$ is not so clear as it is the case for the maximum, c.f. Fig. \ref{fig3} (a) vs (b). For this reason we considered only the localized $\mathsf{v}_0 = 0$ and the first running $\pm \mathsf{v}_1 = \pm \omega$ states. 
E.g., for temperature $Q = 0.1$ the corresponding mean residence time reads $\tau_0 = 509.81, \tau_{\pm 1} = 4.4$. Consequently the locked state is much more stable against thermal fluctuations than the running solutions. It implies a highly counter-intuitive behavior in which the diffusion coefficient is much smaller than the Einstein free diffusion coefficient $D = 0.01 D_0 \ll D_0$, see Fig. \ref{fig2}, despite the fact that temperature is high $Q = 0.1$ and the particle is driven by the external harmonic force with large amplitude $a = 9.81$. When temperature increases the locked state is  destabilized but it is still less sensitive to perturbations than the running ones. E.g. for temperature $Q = 1$ we find $\tau_0 = 10.11, \tau_{\pm 1} = 5.55, \tau_{\pm 2} = 5.02$. An inspection of Fig. \ref{fig4} (d) reveals that even for the high temperature limit frequent segments of localized states in the particle trajectory are interspersed by still rare but long excursions. If temperature is further increased there are more and more unstable short-living periodic orbits which are more frequently visited  and  the spread of trajectories grows, details of deterministic structure is washed out and the diffusion coefficient behaves as the Einstein diffusion coefficient.

\begin{figure}[t]
	\centering
	\includegraphics[width=1.0\linewidth]{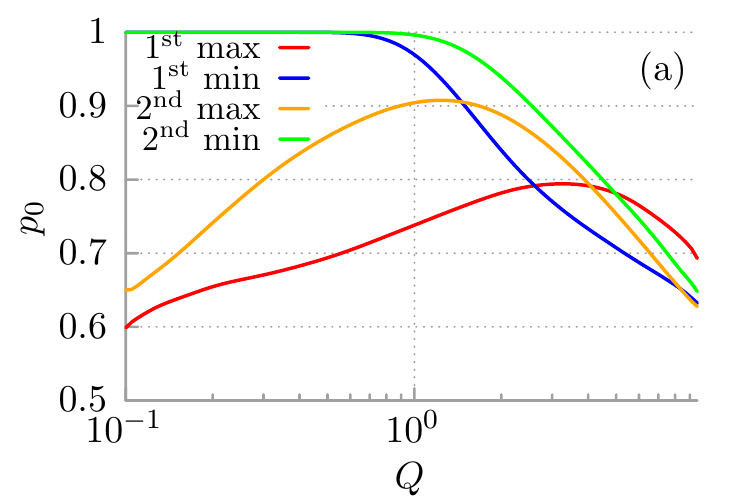}
	\includegraphics[width=1.0\linewidth]{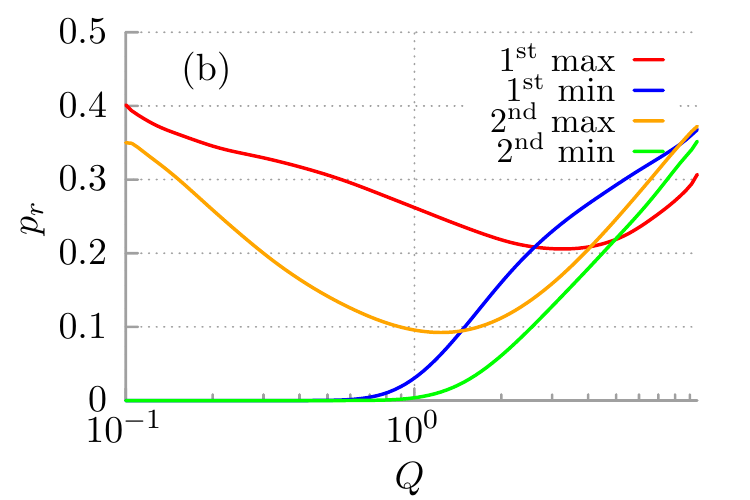}
	\caption{The stationary probabilities to reside in the localized $p_0$ (panel a) and running $p_r$ (panel b) states as a function of temperature $Q$. The driving amplitudes $a$ are fixed to the values where the first two maxima and minima of the diffusion coefficient $D$ are observed. Other parameters are the same as in Fig. \ref{fig2}.}
	\label{fig6}
\end{figure}
\subsection{Residence probabilities}
Since our system is symmetric a preferential direction of motion is impossible in the asymptotic stationary state implying $\langle v \rangle = 0$. This condition imposes a fundamental constraint on the impact of thermal fluctuations onto the set of deterministic solutions for the particle velocity. The latter can be quantified by temperature dependence of the stationary probabilities to reside in each of the states existing in the noiseless case. From a qualitative perspective there are only two distinct classes of states, namely the locked $\overline{v} = 0$ and running $\overline{v} \neq 0$ one. Therefore in Fig. \ref{fig6} we present the stationary probabilities $p_0$ and $p_r = 1 - p_0$ that the particle occupies the locked and running state, respectively, versus temperature of the system. 

We detect that at the minima of the diffusion coefficient  the probability $p_0 = 1$ in a wide interval of temperature $Q \in [0.1,1]$. It means that thermal fluctuations destroy the running solutions and all trajectories eventually are localized in the locked state in which by definition the time average velocity $\overline{v} = 0$ and consequently the symmetry condition is satisfied $\langle v \rangle = 0$. It explains why the diffusion coefficient $D$ at the minima can be reduced even well below its value for a free particle. We note that for asymmetric systems thermal noise induced dynamical localization in a running state can lead to emergence of transient but long-lasting subdiffusion without broad distributions or strong correlations traditionally identified with disorder, trapping, viscoelasticity of the medium or geometrical constraints \cite{spiechowicz2017scirep,spiechowicz2019chaos}. In contrast, here localization in the locked state induces normal diffusion but with an extremely small diffusion coefficient. If temperature increases at the minima the probability to reside in the locked state $p_0$ detaches from one and decreases while at the same time the corresponding quantity for the running state $p_r$ grows. All this causes the diffusion coefficient to increase with temperature as it is commonly expected.

On the other hand, in the symmetric system localization can occur also in a pair of the running states transporting the particle into the opposite directions so that the symmetry condition $\langle v \rangle = 0$ is still satisfied. Since such a type of trajectories maximizes the spread it has the greatest impact on the diffusion coefficient. It can be inferred from the panel (b) of Fig. \ref{fig6} where the probability $p_r$ to reside in the running state is depicted as a function of temperature. It turns out that temperature dependence of $p_r$ fully characterizes the diffusion coefficient $D$ as a function of temperature, compare Fig. 2(b) and Fig. 6(b). When the latter decreases $p_r$ is reduced whereas if $D$ grows $p_r$ increases as well. The positions of extremes at the maxima of diffusion also agree. This correspondence occurs due to the symmetry of the system which imposes the condition of equal coexistence of states with opposite velocities. However, in general the quantifier $W = 1 - |p_0 - p_r|$ which loosely speaking describes the difference between the number of locked and running trajectories could be used to qualitatively predict the behavior of the diffusion coefficient \cite{march-chaos}. We note that in the presently studied case $p_0 > 0.5$ and for this reason $W = 1 - p_0 + p_r = 2 p_r$. Therefore the difference between the number of locked and running solutions is characterized by the probability $p_r$ alone.

\section{Summary}
With this study we numerically analyzed diffusion of the Brownian particle moving in a spatially periodic potential and driven by an unbiased time-periodic force. Our investigation was focused mainly on  temperature dependence of the diffusion coefficient $D$ in the regime in which its quasi-periodicity with respect to the driving amplitude is observed. 

We revealed that for lower temperature the diffusion coefficient at the minima is extremely reduced even well below the Einstein diffusion coefficient for a free particle. The origin of this effect lies in thermal noise induced localization of all system trajectories in the locked state in which the particle motion is bounded to a finite number of the periodic potential wells. If temperature grows the diffusion coefficient increases as it is commonly expected. In contrast, for the same lower temperature this quantity at the maxima is extremely large. If temperature increases in this regime the anomalous behavior is detected in which the diffusion coefficient decreases when temperature grows. As a reason for this intriguing effect we identified peculiar influence of thermal fluctuations on transitions between two classes of trajectories, namely the locked and running ones, existing in the deterministic dynamics of the system. In particular, we found out that temperature dependence of the diffusion coefficient in its minima and maxima can be qualitatively reconstructed by temperature dependence of the stationary probability to reside in the running state. 

As a final note we point out that it is not a nuisance that the non-monotonic temperature  behavior of diffusion coefficient is related to proportion between the number of locked and running trajectories but seemingly it emerges rather as an universal property observed for different systems in various regimes \cite{speer2012, spiechowicz2017scirep, spiechowicz2017chaos, marchenko2018, lind}.
 
\section*{Acknowledgments}
This work was supported by the Polish Grants NCN: 2022/45/B/ST3/02619 (J.S.), 2018/30/E/ST3/00428 (I. G. M.) and in part by PL-Grid Infrastructure. V.A. is supported by the Ukrainian Grant Agency NASU No. 0122U002145/2022-2023. I. G. M. acknowledges University of Silesia for its hospitality  since the beginning of the war, 24 February 2022.  

\appendix
\section{Scaling of the Langevin equation}
We consider a classical Brownian particle of mass $M$ subjected to a one-dimensional, spatially periodic potential $U(x)$ and driven by an unbiased and symmetric time-periodic force $F(t)$. Its dynamics can be described by the Langevin equation in the form \cite{entropy23}
\begin{equation}
	\label{model1}
	M\ddot{x} + \Gamma\dot{x} = -U'(x) + F(t) + \sqrt{2\Gamma k_B T}\,\xi(t),
\end{equation}
where the dot and the prime denotes  differentiation with respect to time $t$ and the particle coordinate $x$, respectively. The parameter $\Gamma$ is the friction coefficient. The potential $U(x)$ is assumed to be symmetric of spatial period $L$ and the barrier height  $2 \Delta U$ reading
\begin{equation}
	\label{potential}
	U(x) = U(x+L) = -\Delta U\cos{\left( \frac{2\pi}{L}x \right)}.
\end{equation}
The external ac-driving force of amplitude $A$ and angular frequency $\Omega$ has the simplest harmonic form  
\begin{equation}
	F(t) = A \sin{(\Omega t)}. 
\end{equation}
Thermal equilibrium fluctuations due to interaction of the particle with its environment of temperature $T$ are modelled as $\delta$-correlated Gaussian white noise of zero-mean value, 
\begin{equation}
	\langle \xi(t) \rangle = 0, \quad \langle \xi(t)\xi(s) \rangle = \delta(t - s), 
\end{equation}
where the bracket $\langle \cdot \rangle$ denotes an average over white noise realizations (ensemble average). 
The noise intensity $2\Gamma k_B T$ in Eq. (\ref{model1}) follows from the fluctuation-dissipation theorem \cite{kubo1966}, where $k_B$ is the Boltzmann constant. If $A=0$ the stationary state is a thermal equilibrium state. If $A \neq 0$, then the external force $F(t)$ drives the system away from the equilibrium state. 

Now, we transform Eq. (\ref{model1}) to its dimensionless form. To this aim we use the following scales as  characteristic units of length and time
\begin{equation}
	\label{scaling}
	\hat{x} = 2\pi \frac{x}{L}, \quad \hat{t} = \frac{t}{\tau_0}, \quad \tau_0 = \frac{L}{2\pi}\sqrt{\frac{M}{\Delta U}}.
\end{equation}
Under such a procedure Eq. (\ref{model1}) assumes the form
\begin{equation}
	\label{dimless}
	\ddot{\hat{x}} + \gamma\dot{\hat{x}} = -\sin{\hat{x}} + a \sin (\omega \hat{t}) +  \sqrt{2\gamma Q} \, \hat{\xi}(\hat{t}).
\end{equation}
The  dimensionless parameters  are defined by the relations 
\begin{eqnarray}
	\gamma = \frac{\tau_0}{\tau_1}, \quad a = \frac{1}{2\pi}\frac{L}{\Delta U} A,  \nonumber\\
\quad \omega = \tau_0 \Omega, \quad Q = \frac{k_B T}{\Delta U}, 
\end{eqnarray}
where the second characteristic time  is $\tau_1 = M/\Gamma$. It has the physical interpretation of the relaxation time for the velocity of the free Brownian particle. On the other hand, the characteristic time  $\tau_0$ is related to the period of small  oscillations inside the potential $U(x)$ wells. 

The rescaled potential $V(x)$ of the period $L=2\pi$ reads $V(\hat{x}) = U((L/2\pi)\hat{x})/\Delta U = -\cos{\hat x}$ and the corresponding potential force is $f(x) = -\hat{U}'(\hat{x})=-\sin \hat{x}$. The rescaled thermal noise is $\hat{\xi}(\hat{t}) = (L/2\pi \Delta U)\xi(t) = (L/2\pi \Delta U)\xi(\tau_0\hat{t})$ and has the same statistical properties as $\xi(t)$; i.e., $\langle \hat{\xi}(\hat{t}) \rangle = 0$ and $\langle \hat{\xi}(\hat{t})\hat{\xi}(\hat{s}) \rangle = \delta(\hat{t} - \hat{s})$. The dimensionless noise strength  $Q$ is the ratio of thermal energy and half of the activation energy the particle needs to overcome the nonrescaled potential barrier.  In order to simplify the notation  we omit the hat-notation in Eq. (\ref{dimless-model}). \\

\section{Description of the simulations}
The complexity of stochastic dynamics determined by Eq. (\ref{dimless-model}) with three-dimensional phase space  $\{x, y=\dot x, z=\omega t\}$ is rooted in the four-dimensional parameter space $\{\gamma, a, \omega, Q\}$.  The  Fokker-Planck equation corresponding to Eq. (\ref{dimless-model}) cannot be solved analytically and for this reason we had to resort to comprehensive numerical simulations. All calculations have been done using a Compute Unified Device Architecture (CUDA) environment implemented on a modern desktop Graphics Processing Unit (GPU). This proceeding allowed for a speedup of factor of the order $10^3$ times as compared to present day Central Processing Unit (CPU) method \cite{spiechowicz2015cpc}. The Langevin equation (\ref{dimless-model}) were integrated using a second order predictor-corrector scheme \cite{platen} with the time step $h = 10^{-2}$. The quantities characterizing diffusive behaviour of the system were averaged over the ensemble up to $N = 2^{20} = 1048576$ trajectories, each starting with different initial conditions $x(0)$, $v(0)$ and $\phi$ distributed uniformly over the intervals $[0, 2\pi]$, $[-2,2]$ and $[0, 2\pi]$, respectively. %It implies that the statistical error of the Monte Carlo simulation is estimated as $1/\sqrt{N} \approx 0.008$ which is more than adequate for our purposes. 
The time span of simulations read $10^6$ periods $2\pi/\omega$ of the external driving $a\cos{(\omega t)}$ and was extensive enough to reach the long time limit indicated by the stationarity of diffusion coefficient.

\end{document}